\documentclass[aps,prl,twocolumn,groupedaddress]{revtex4-2}
\usepackage{blindtext}
\usepackage{graphicx}
\usepackage{hyperref}
\usepackage{tabularx}
\usepackage{subcaption}
\usepackage{caption}

\hypersetup{
    colorlinks=true,
    linkcolor=blue,
    filecolor=magenta,      
    urlcolor=cyan,
    pdftitle={Overleaf Example},
    pdfpagemode=FullScreen,
    citecolor=black
    }

\begin{document}
\title {SmartControllerJS: A JavaScript library to turn smartphones into controllers for web-based interactive experiments}

\author{Emma Poliakova}
\email[]{emmka.poliakova@gmail.com}

\author{Fraser Dempster}
\email[]{fraserdempster23@gmail.com}

\author{Abubakr Mahmood}
\email[]{abubakrmahmood@hotmail.co.uk}

\author{Jonathan Grizou}
\email[]{jonathan.grizou@glasgow.ac.uk}
\homepage[]{jgrizou.com}
\affiliation{School of Computing Science, University of Glasgow}


\date{\today}

\begin{abstract}
We introduce SmartControllerJS, a new JavaScript library for fast, cost-effective designing of web applications controlled via everyday smartphones. At its core, SmartControllerJS establishes a connection between two webpages, one page running on a desktop browser and the other on the user's smartphone. The smartphone webpage loads a controller interface allowing users to control a web application running on their computer’s browser. The SmartControllerJS framework enables fast iteration loops when designing interactive user experiments because it has minimal friction and allows for scaling, while having no running costs. We first describe how this library is built, how it can be used, and provide interactive examples. We then present two games designed for public screens along with results from user studies evaluating acceptability and ease of use. Finally, we implement a custom controller based on user feedback and introduce connection monitoring tools. We believe SmartControllerJS can accelerate the design of interactive experiments for researchers in Human-Computer Interaction, and be a useful tool for educational projects. All our code is available at \href{https://github.com/SmartControllerJS}{SmartControllerJS} and links to all demos can be found in Table \ref{table:tableofdemos}. To explore our demos, we recommend reading this work on a desktop computer with your smartphone in hand.

\end{abstract}


\maketitle

\section{Introduction}

Smartphones are ubiquitous and incredibly optimised devices. Virtually everyone possesses a smartphone in today's society, allowing them access to the internet and various sensor functionalities when needed. Features such as phone calling and social media merely scratch the surface of what these portable devices are capable of, and many researchers in the Human-Computer Interaction (HCI) community have explored more innovative uses of smartphones to enhance interactions with the world around us \cite{lucero_holopainen_jokela_2022,alireza,jurgen}. To our knowledge, there is a lack of libraries or frameworks that enable researchers to quickly and efficiently design smartphone-based interactive experiences. We developed SmartControllerJS with the goal of addressing this gap in mind. \\
SmartControllerJS aims to harness the large array of sensors smartphones offer to create controllers that do not require expensive equipment or software installation to operate. The core concept is to turn a user's smartphone into a controller - such as a joystick, a touchpad, or a sensor - simply by scanning a QR code. \\

With this in mind, SmartControllerJS was designed to:

\begin{figure}[b]
    \centering
    \includegraphics[width=0.85\linewidth]{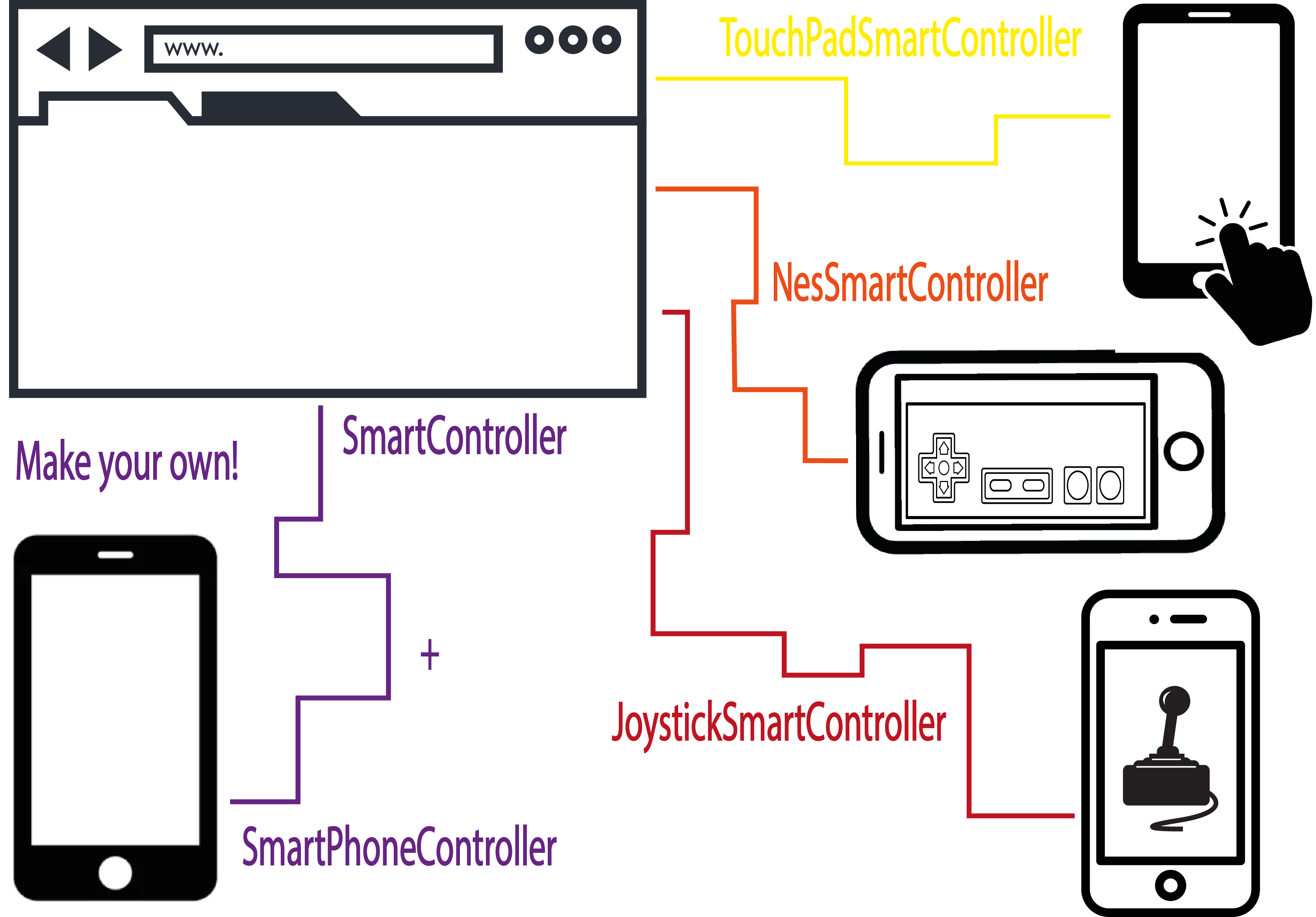}
    \caption{Users can choose to apply any of our ready-to-use controllers, or design their own.}
    \label{all-controllers}
\end{figure}    

\begin{enumerate}
\item \textbf{Achieve minimal friction for both developers and users.} As a developer, the library abstracts all connection and communication aspects and provides base classes for developers to build on; only two lines of code are necessary to get started. As a user, our standalone static webpage approach ensures that no installation will be required.
\item \textbf{Cost nothing, even with large-scale implementations.} We rely on peer-to-peer communication between devices using the WebRTC protocol, bypassing the need for servers to handle communication, even at larger scales. 
\item \textbf{Allow ease of sharing.} Webpages and link sharing are familiar concepts to most people.
\item \textbf{Allow multiple users to connect to a single display.} Enabling different multiplayer implementations to be explored using the library. This can be especially useful in public scenarios.
\item \textbf{Allow the monitoring of the connection statistics.} This would enable tracking of the impact that message rates have on user experience.
\item \textbf{Be widely accessible.} Our work is open-source, hosted publicly on \href{https://github.com/SmartControllerJS}{GitHub} with packages released on package managers  \href{https://www.npmjs.com/package/smartcontroller}{NPM} and \href{https://unpkg.com/browse/smartcontroller@3.2.7/}{UNPKG}.

\end{enumerate}

\section{SmartControllerJS}

In this section we explain how the individual components of SmartControllerJS fit together to provide a robust infrastructure for quick and easy creation of controllers and matching web applications. A detailed list of all the demos can be found in table \ref{table:tableofdemos}.

\subsection{Overview of library}

\begin{figure*}[b]
\includegraphics[width=0.66\linewidth]{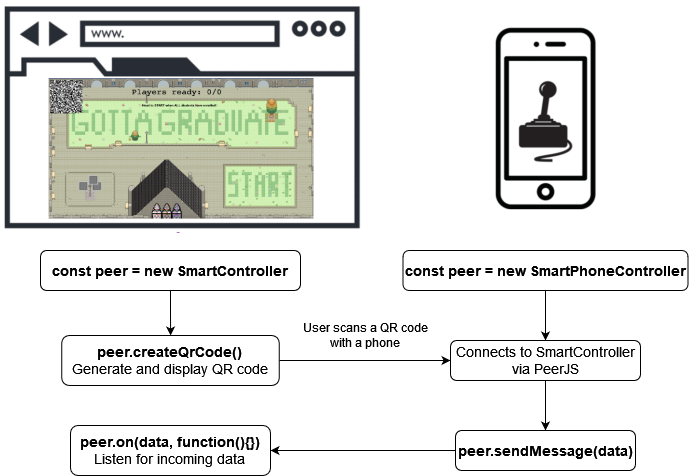}
\caption{\label{scheme}  Flow chart of basic steps to establish the peer-to-peer connection and send data.}
\end{figure*}

\href{https://github.com/SmartControllerJS/SmartController}{SmartControllerJS} is written in JavaScript. It uses the \href{https://peerjs.com/}{PeerJS} library to facilitate the peer-to-peer connection, and \href{https://github.com/EventEmitter2/EventEmitter}{EventEmitter2} to monitor and trigger events. The library is bundled with Parcel and published on \href{https://www.npmjs.com/package/smartcontroller}{npm}, the most popular package manager for JavaScript libraries and widely used by web developers. The two main building blocks of the library are the SmartController and SmartPhoneController classes. Figure \ref{scheme} gives the general overview of how the two classes work together. \\

The SmartController class is used in the computer browser, managing all incoming connections from smartphones and handling the data input. The class also creates and handles various events, such as new connections, incoming data, and closing of channels. The PeerJS object instantiates a peer ID to which a listener can establish a connection. The SmartController class maintains a list of connected smartphones listed by peer IDs and creates controller objects that store and manage the data of each smartphone. The class is also responsible for deleting list entries when a user disconnects. Lastly, the SmartController has a method for creating and displaying QR codes, using \href{https://github.com/ushelp/EasyQRCodeJS}{EasyQRCodeJS}. \\

SmartPhoneController is a class managing the peer connection and message passing on the smartphone browser. The smartphone-to-computer connection is fully automated and handled upon SmartPhoneController object creation, and all parameters necessary for the connection are specified and read from the URL. The SmartPhoneController class has a single method to send data to the computer browser, which can include statistics or any relevant information produced by the users’ interaction with the controller. The message frequency can be limited by setting a throttle that specifies how often messages should be sent during an interval, ignoring messages produced outside the specified interval.

\subsection{Connection}

An integral aspect of SmartControllerJS is the use of Web Real-Time Communication or WebRTC.  \href{https://webrtc.org/}{WebRTC} allows establishing peer-to-peer (P2P) communication between browsers without requiring an intermediary server or application and allows sharing video, audio, and generic data.  It only requires a lightweight signaling server for the initial set-up by allowing two peers to “find each other". This server is free of charge and PeerJS allows the users to deploy their own instance should they wish to do so.  \\

\onecolumngrid

\clearpage
\twocolumngrid

In the SmartController library, a peer ID identifies the computer browser and is passed as an argument to the URL of the smartphone controller. For example, for the touchpad controller, an application with a peer ID of ‘123456789’ should ask the user to open \emph{https://smartcontrollerjs.github.io/Controllers/touchpad .html?id=123456789} on their smartphone. A QR code is automatically generated to pass this information to a smartphone with minimal friction.\\

Once a connection between a smartphone and a computer is established, the SmartController will store this connection in a list before generating and storing the new controller object. There is no limit on the number of connections that can be established. For identifying specific users, an argument, called playerid, can be passed to the controller URL to uniquely identify a user. Following our example above, a player called JohnDoe, could be identified as \emph{https://smartcontrollerjs.github.io/Controllers/touchpad .html?id=123456789\&playerid=JohnDoe}.\\ 

\begin{figure}[t]
    \centering
    \includegraphics[width=\linewidth]{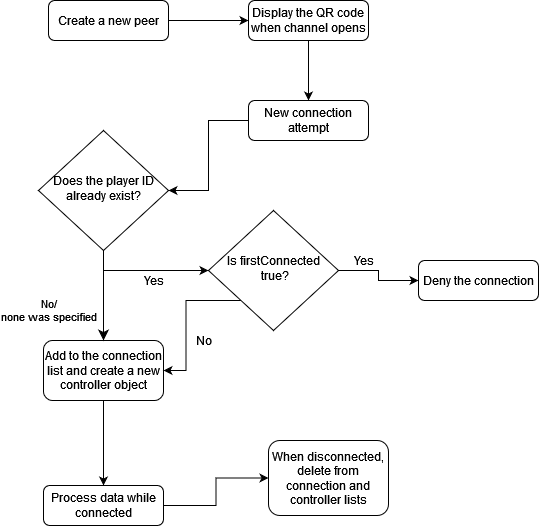}
    \caption{\label{state} Diagram defining the state upon peer creation.}
    \label{statediagram}
\end{figure}

Being able to set a player ID is helpful for multiplayer games, such that a user can be assigned a specific role. We include an optional parameter, called firstConnected, to limit connection for a player ID to the first person connecting, as shown in Figure \ref{statediagram}. This prevents multiple users having the same player ID and causing errors. If there is a player ID specified and firstConnected is true, SmartController will check if an ID is already stored in the connection list. If yes, no further connections to that ID will be allowed. If firstConnected is set to false, every new connection to the same player ID will overwrite the original one.

\subsection{Controllers}

The foundation of each controller type is a BaseController class, which is responsible for processing data for individual smartphones. As this is the most basic, default version of a controller object, it only stores a peer ID, a player ID (if specified), and statistic fields for ping and message rate. For simplicity, each new controller is created by extending this class, adding new fields and modifying the base method to expand the functionality. \\

In the smartphone browser, controllers only need to use the SmartPhoneController class to manage the connection and send data to the computer. The community has already created a range of controllers hosted on \href{https://github.com/SmartControllerJS/Controllers}{GitHub} that can be reused in other browser application. Current controllers include an NES controller, a joystick, a touchpad, and an accelerometer. \\

Each controller has a corresponding class that extends BaseController and provides additional fields and methods specific to the controller. For example, the TouchPad class adds a boolean isActive state that is True when the player is interacting with the phone screen and False otherwise, and the NesController class stores a button dictionary containing the states of all buttons.

\subsection{Statistics}

\begin{figure}[b]
    \centering
    \includegraphics[width=\linewidth]{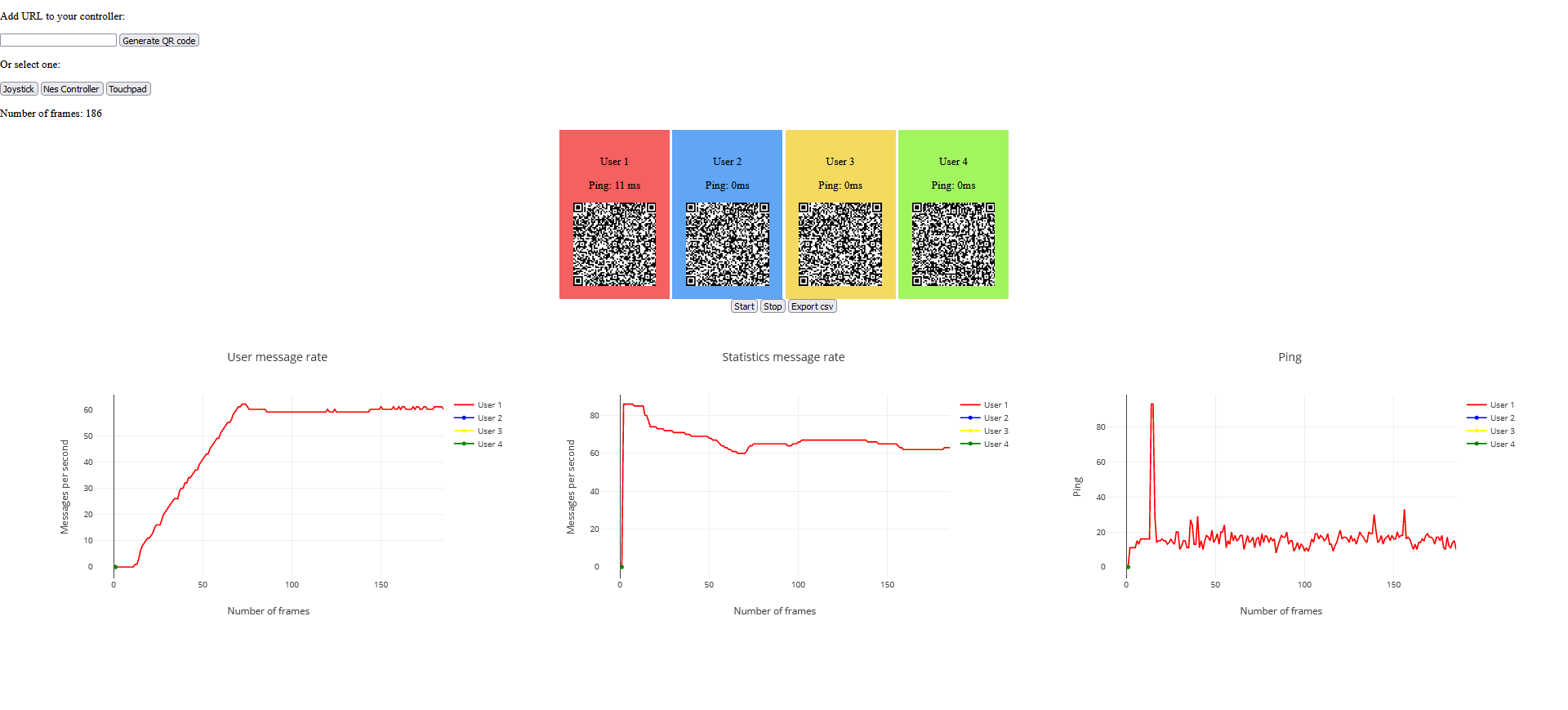}
    \caption{Statistics website used to measure ping, user and statistics message rates for any controller created with SmartController.}
    \label{stats}
\end{figure}

There is a range of different statistics available to the user to monitor the performance of the library. These can be switched on and off as needed. If the statistics parameter is switched off then ping and the statistics message rate are not available. A dedicated web page for measuring ping, user message rate, and statistics message rate was created to help the users test various controllers.
The first statistic available is ping. It is calculated in milliseconds as a round trip between a computer and a smartphone and stored as a dedicated field for each controller. Using the device locally resulted in a ping of around 20 ms, while the remote devices oscillated between 40 ms and 80 ms in the case of the joystick and 40 ms - 60 ms for the touchpad.
The other statistic is message rate, which is calculated separately for user-type messages and statistic-type messages. It captures the frequency of messages sent from the smartphone to the computer. Unlike with ping, the higher numbers are desirable here. When the rate is high it means we can achieve greater accuracy in control. The message rate was in general measured to be very high, ranging from 60 messages to 150 messages per second.\\

Figure \ref{stats} shows a screenshot of \href{https://smartcontrollerjs.github.io/Controllers/stats.html}{statistics webpage} used for measuring ping, user message rate, and statistics message rate. The user can input a URL to their own controller or use one of the available ones to measure performance in different areas. The webpage can export the data from the evaluation to CSV. The graphs are created with the \href{https://plotly.com/}{Plotly} library. It offers many convenient plotting functions, one of which is the ability to directly download the created graphs.

\section{Interactive Demos}
\begin{figure*}[b]
        \centering
        \begin{subfigure}[b]{0.475\textwidth} 
            \centering
            \includegraphics[width=\linewidth]{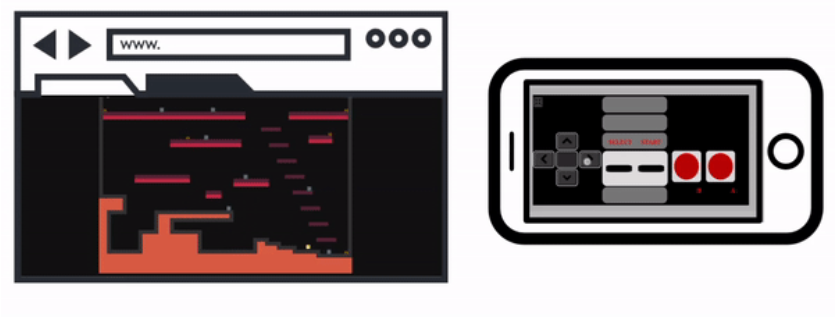}
             \caption[Platform game with an NES controller. ]%
            {{\small Platform game with an NES controller.}} 
            \label{platform}
        \end{subfigure}
        \hfill
        \begin{subfigure}[b]{0.475\textwidth}  
            \centering 
            \includegraphics[width=\textwidth]{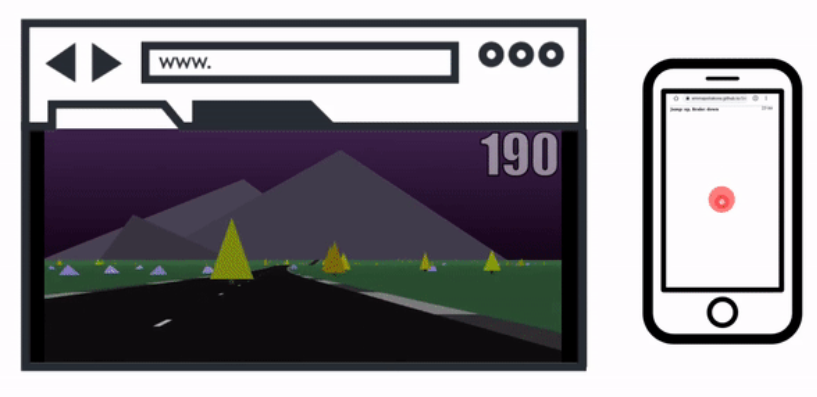}
            \caption[3D racing game with a joystick.]%
            {{\small 3D racing game with a joystick.}}    
            \label{racer}
        \end{subfigure}
        \vskip\baselineskip
        \begin{subfigure}[b]{0.475\textwidth}   
            \centering 
            \includegraphics[width=\textwidth]{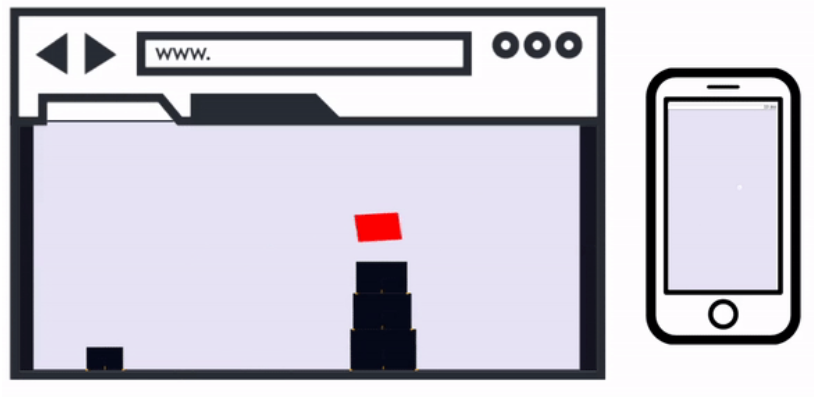}
            \caption[Single player physics simulator controller with a touch-pad.]%
            {{\small Single player physics simulator controller with a touch-pad.}}    
            \label{physics}
        \end{subfigure}
        \hfill
        \begin{subfigure}[b]{0.475\textwidth}   
            \centering 
            \includegraphics[width=\textwidth]{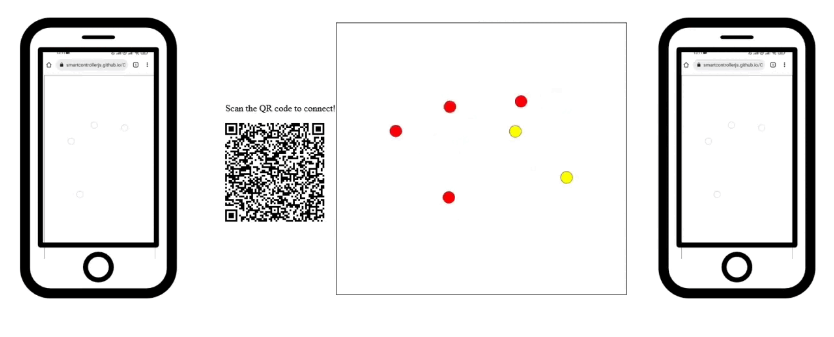}
            \caption[Multiplayer touch-pad demo used as a simple tutorial.]%
            {{\small Multiplayer touch-pad demo used as a simple tutorial.}}    
            \label{touchpad}
        \end{subfigure}
        \caption[ A demonstration of the games transformed to be used with SmartControllerJS. ]
        {\small A demonstration of the games transformed to be used with SmartControllerJS.} 
        \label{demos}
    \end{figure*}

Figure \ref{platform} shows a simple \href{https://emmapoliakova.github.io/WebRTCSmartphoneController/demo/tinyPlatformer/index.html}{2D platformer}, where the player needs to collect all the coins and avoid the enemies, re-made to use our NES SmartController. 

The \href{https://emmapoliakova.github.io/WebRTCSmartphoneController/demo/3dRacing.html}{3D racing game} was originally controlled with a computer mouse. It can now be played with a joystick, as shown in Figure \ref{racer}.  

The \href{https://emmapoliakova.github.io/WebRTCSmartphoneController/physics/physicsDemoV3.html}{physics simulator} in Figure \ref{physics} is controlled via a touch-pad and lets the user move blocks around on the screen and stack them. This can also be played as \href{https://emmapoliakova.github.io/WebRTCSmartphoneController/physics/physicsDemoV4.html}{multiplayer}.  

Finally, figure \ref{touchpad} demonstrates a simple \href{https://smartcontrollerjs.github.io/Controllers/touchpad-receive.html}{multiplayer demo} created as a guide to let developers learn SmartControllerJS. 

All demos can also be accessed via \href{https://smartcontrollerjs.github.io/SmartController/}{the documentation page}. 

\onecolumngrid

\clearpage
\twocolumngrid

\newpage
\section{Interactive Experiments}

The following sections focus on two projects which incorporate SmartControllerJS to create multiplayer games to explore the capabilites of the library. They each contain a description of the project work, and developer and user feedback.

\section{Experiment 1 - Gotta Graduate}

Gotta Graduate is a multiplayer cooperative game made with \href{https://phaser.io/}{Phaser}. It takes advantage of the SmartController concept and the Joystick Controller to provide a more accessible experience for users whilst also allowing the game to be played in public spaces.\\

\begin{figure}[b]
    \centering
    \includegraphics[width=\linewidth]{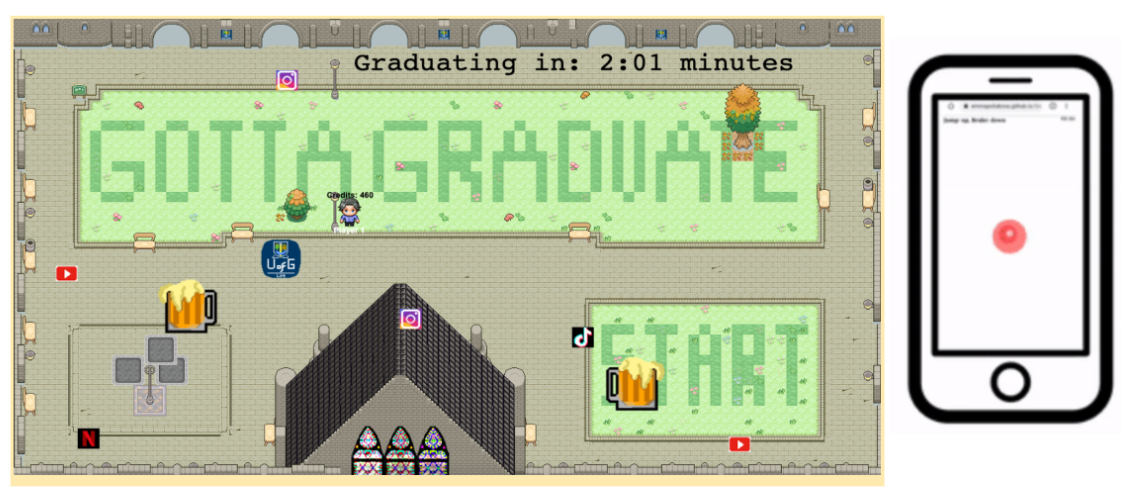}
    \caption{Gotta Graduate with joystick controller}
    \label{stategotta}
\end{figure}

You can play the game online \href{https://fraser-dempster.github.io/l4-project-interactive-game/}{here} and read more about the gameplay and motivation at \href{https://github.com/SmartControllerJS/GottaGraduate}{Gotta Graduate}. In the game, you take the role of a student that can walk around campus. Your goal is to graduate by avoiding distractions such as YouTube, Netflix, or TikTok and grabbing University credits. It is multiplayer and the game will start when all players reach the START area in the bottom-right corner.\\

Gotta Graduate was used to evaluate the effectiveness of the SmartController library. An experiment was conducted with 32 participants from the university and the public. The experiment consisted of five stages: a pre-study survey, playing Gotta Graduate, a post-study survey, observations (where possible), and interviews (on participant agreement).\\

Before starting the game, pre-study questions gathered data on users’ demographic, comfort with launching QR codes, experience using interactive displays, and prior gaming experience. Demographic data was used to identify discrepancies among participant understanding of the SmartController. \\

After the pre-study questions were completed, participants were prompted to play one round of Gotta Graduate lasting two minutes and thirty seconds. To keep findings consistent across participants during the experiment, no extra information was given aside from a brief explanation of what Gotta Graduate was and instructions on how to complete the survey. This allowed us to see how a user would naturally interact with Gotta Graduate using the SmartController and how intuitive they found the experience.\\

Once participants played a round of Gotta Graduate, they completed a post-study survey. The survey contained two sections: usability testing and feelings invoked. Both sections presented participants with statements which they could respond to using a Likert scale ranging from strongly agree (1) to strongly disagree (5). These two sections evaluated the significance of the gameplay, impact of the design choices, the implementation techniques, and technologies used. Within the usability section, 
SmartController-relevant statements included “The game was easy to connect to using the QR code”, “Starting the game took little effort”, and “The controller was intuitive”. Statements from the second section included “I felt I was engaged with the game”.\\

When possible, observations were made while participants played Gotta Graduate. How quickly players started playing and how intuitive they found the controller was of particular interest.\\

\begin{figure}[b]
    \centering
    \includegraphics[width=\linewidth]{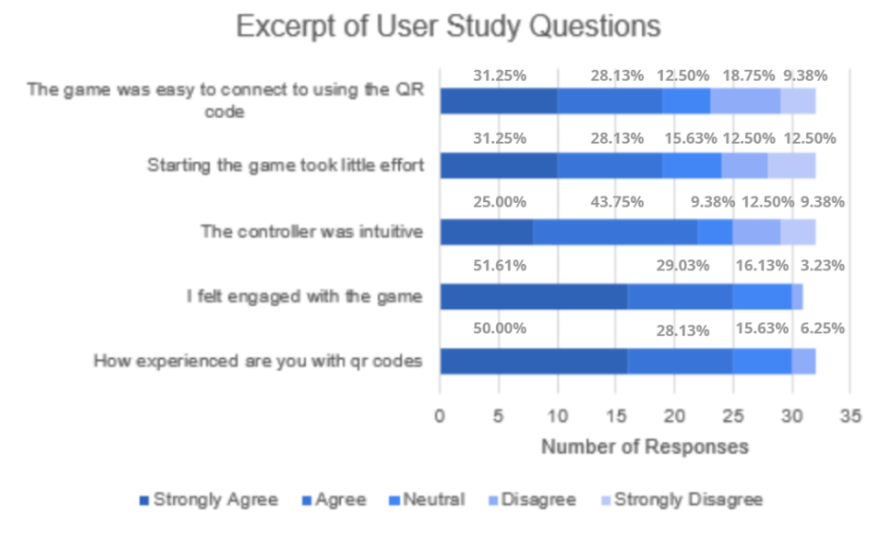}
    \caption{A plot summarising the relevant results from the conducted survey}
    \label{stategottaresult}
\end{figure}

Lastly, if the participant agreed, interviews were carried out after the session. Interviews followed a semi-structured style and began with an open-ended question: “How would you improve interaction between players?”. The open-ended nature of the questions led to many avenues of conversations including the intuitiveness of the SmartController and any learning curves present.\\

In total, 32 participants took part in the experiment. The results relevant to the SmartController are summarised in Figure \ref{state}.

\subsection{Pre-study results}
84.4\% of participants were aged 18-28 years old and marked their occupation as “student”. When asked how experienced they were with QR codes, 100\% of participants had experience with 50\% being very experienced. 

\subsection{Post-study Results}
It was found that 69.8\% agree or strongly agreed that using the SmartController was intuitive, whilst 21.9\% disagreed. Of the 32 participants, 59.4\% agreed or strongly agreed that starting the game took little effort while 25\% disagreed or strongly disagreed. When asked if the game was easy to connect to using the QR code, 59.4\% agreed or strongly agreed and 28.1\% disagreed or strongly disagreed. Finally, an
overwhelming 78.1\% of participants agreed or strongly agreed that the game was engaging while one participant disagreed. It should be noted that one participant failed to respond to that question.

\subsection{Observations} 
Out of the 32 participants, ten were observed while playing one round of Gotta Graduate. Four participants struggled to understand how to use the controller when first attempting to play whereas six participants immediately began playing with it. Groups of participants playing together were seen helping each other understand how to use the controller which improved learning times.

\subsection{Interviews}
Three participants agreed to be interviewed after completing the experiment. When asked whether “they would show friends the game”, one participant mentioned that the game was fairly easy to start playing and set up which would mean that their friends would understand the game easily so it would not be embarrassing to introduce them to it.\\

In a real life situation, it is clear that the SmartController is an effective means of having users interact with an application using their smartphone. Participants found the game easy to connect to using the SmartControllers’ QR code. In addition, users found the controller intuitive and quickly learnt how to play Gotta Graduate. The low barrier of entry ensures that applications are accessible and support a wide audience of users. It should also be noted that most participants were already experienced with using QR codes before using the SmartController. The results consolidated our aims and established the SmartController’s prowess as an alternative to typical methods of interaction.

\section{Experiment 2 - Coin Chaser}

Coin Chaser is a 2D multiplayer game that aims to take on the role of a public interactive experience, while simultaneously demonstrating the use of the SmartController library with Unity, the industry-standard game engine. \\

\begin{figure}[b]
    \centering
    \includegraphics[width=\linewidth]{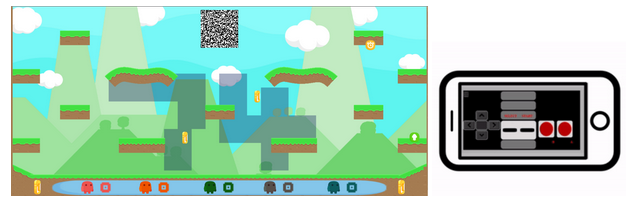}
    \caption{The level design of Coin Chaser with the chosen NES controller.}
    \label{oldcontroller}
\end{figure}

The players are each tasked with collecting coins to add to their individual scores while also collecting various character effects that would influence their style of play. Upon scanning the displayed QR code (generated using SmartController) with their smartphone, the user is redirected to a webpage presenting the NES controller configuration seen in Figure \ref{oldcontroller}. Coin Chaser awakens a 2D player character on the display when a participant scans the QR code, and then allows the user to control the character through the use of horizontal movement and the ‘A’ button on the smartphone controller. Five players are able to interact with Coin Chaser simultaneously, achieving multiplayer functionality. The game can be played from \href{https://smartcontrollerjs.github.io/Coin-Chaser/}{here} and more information on its development can be accessed at \href{https://github.com/SmartControllerJS/Coin-Chaser}{Coin Chaser}.\\

The main challenge in developing Coin Chaser was that the SmartController package was developed using JavaScript whereas the game was implemented using Unity with C\# scripts. Therefore, it was necessary that we enable communication between the browser script and Unity game instance to share input data. \textit{WebGL}, which was used to render the game on a browser host, has a function that enables communication from a website browser script to the rendered game on a webpage.  This function, \href{https://docs.unity3d.com/Manual/webgl-interactingwithbrowserscripting.html}{\textit{SendMessage}}, relays inputs from the browser host to the rendered game, where the inputs have been sent from the NES SmartController to the browser host.\\

The evaluation for Coin Chaser was conducted in a controlled environment, using a computer as the display, and by asking groups with a varying number of participants to interact with the game on the screen. Groups of participants were given a set task to interact with the display, rather than monitoring interaction from random passersby with the game as we would in a public space.  \\

The study for the evaluation was designed in three stages as follows:
\begin{enumerate}
    \item A pre-interaction questionnaire took place prior to asking participants to interact with Coin Chaser. This section of the evaluation aimed to assess both the pre-existing gaming and QR code scanning experiences.
    \item An in-game interactions of participant groups were then evaluated during their interaction experiences with Coin Chaser using the Think-Aloud technique. The Think-aloud technique requires participants to narrate their thoughts to the researcher while completing a set task \citep{charters_use_2003}. To evaluate participant groups as if they were interacting with the display in an imagined public space, we navigated to the game and asked participants to interact with the game while also narrating their thoughts throughout the process of initial smartphone connection establishment, gameplay experience, and eventual controller disconnection.
    \item A post-interaction study which took place after the in-game interaction was aimed to assess participant responses to the experience, and preferences between a real controller, such as an Xbox One controller, or the digital smartphone controller they were given.

\end{enumerate}

Participants were evaluated in groups of sizes varying between one and four participants. Twenty-one participants were successfully recruited for the study, and their responses were recorded accordingly. This total was split into seven single participants, two groups consisting of two participants, two groups consisting of three participants, and one group consisting of four participants. The researcher also joined certain group in-game interaction evaluations to emphasise the possible competitive nature evoked when experiencing the game with multiple players.

\subsection{Pre-interaction Study Results}
The majority of participants claimed to have some gaming experience with only two participants having very little experience, one of which had no experience using a game controller. 100\% of participants had experience in scanning QR codes. 

\subsection{In-game interaction Observations}
The Think-aloud evaluation revealed that:
\begin{itemize}
    \item Participants with more gaming experience adapted quicker to the game controls than those with less experience.
    \item iPhone users had irregular issues with the SmartController package during the evaluation. Most participants with iPhones could not establish a connection. When a connection was made successfully, participants experienced issues with the browser controller interface. Participants found that buttons that were held down were being highlighted and unpressed by the phone, and that pressing two buttons on the controller simultaneously causes the phone to zoom in on the interface.
    \item Participants initially struggled to navigate the controller interface while looking at the game display, however, all participants overcame this by their second game round.
    \item No players encountered issues with interaction or Unity game when using an Android phone.
    \item When using an Android phone, all participants scanned the QR code and established a connection easily without the need for further instruction.
\end{itemize}

\subsection{Post-interaction Study Results}
When asked to score the smartphone controller in comparison to other game controllers using a Likert scale from one to five, one being "Much worse" and five being "Much better", most participants stated that the smartphone controller presented a similar or worse experience. Only two participants gave a score of four on the scale, meaning that they thought the SmartController performed better than other game controllers the participant had experience with. However, no participant thought the smartphone controller was much better than other controller options. Following this question, participants were asked why they prefer one controller over the other. Notable responses included:
\begin{itemize}
\item Physical controllers provide haptic feedback, such as vibrations when pressing a button, whereas our controller does not.
\item Participants using iPhone devices experienced slow response times after pressing a button on the smartphone controller. 
\item Participants using iPhone devices unintentionally highlighted the button on the controller interface rather than selecting it.
\item Physical controllers offer tactile buttons that are easier to navigate without needing to look at the device.
\item The smartphone controller buttons were too small, making participants miss button presses.
\item The smartphone controller was unfamiliar to participants in contrast to the widely used physical controllers.
\item Participants using iPhone devices unintentionally zoomed in on the phone controller interface when two buttons were pressed at the same time.
\end{itemize} 

Participants commented that the controller had unnecessary and unused buttons that take up screen space that could be used more efficiently. One participant suggested the implementation of haptic feedback. There were varying responses on the complexity of the controller interface, with some respondents stating that the controller is simple to use while others respondents believe the opposite to be true.\\

\subsection{Result Analysis}
Overall, we have found that there is potential for the SmartController package to be used with Unity to create an interactive display intended for a public space. After monitoring participant interaction with Coin Chaser during the In-game Interaction Evaluation section, and by collecting comments made on the game in the post-interaction questionnaire, we can see that the overall review for the controller has been positive where an Android phone was used. The noted habits and occurrences from the in-game evaluation show that participants using Android devices faced no real difficulties when interacting with the display using the SmartController. Participants were able to easily scan the displayed QR code above the game instance to establish a connection and all participants were able to adapt to the controller and gameplay before the second game round ended. \\

However, the experience still has room for improvement. iPhone users often failed to establish a connection to the display after scanning the QR code. Even if iPhone users had connected to the game, they experienced issues such as the highlighting of controller buttons and the unintended zooming into the interface. These issues were not observed on Android devices. Therefore, further work must be done to improve the SmartController package to overcome this iPhone-Android device bias to make the display more accessible to a larger fraction of people. \\

Additionally, most participants preferred real controllers to digital smartphone controller. Some responses for why participants in our study held this preference were also experienced by  \citet{baldauf_investigating_2015}. In contrast to physical controllers, digital controllers lack tactile and haptic feedback and cause users to lose track of the buttons they were pressing, causing them to look away from the game and back to the phone to reposition themselves. Despite some lack of familiarity with using smartphones as controllers, the study showed that SmartController is suitable for creating games and demos. From the post-interaction study we can see that the excess of unused buttons proved to be a detriment to the experience as they took up screen space and participants often pressed buttons that they did not intend to press. To resolve some of these issues, a tailored controller for Coin Chaser has now been implemented. From this, we learned that keeping the design of each controller as simple as possible improves user experience.

\subsection{The Tailored Controller}
To overcome the issues that participants experienced with the NES controller, we decided to implement our own tailored controller that is unique to Coin Chaser. This controller aims to improve usability by reducing the number of buttons on the display to in order to mitigate unintentional button presses, and to minimise the need to look from the game to the controller while also retaining the core game control functionalities.\\

To create this controller we added a new \href{https://github.com/SmartControllerJS/Coin-Chaser-Tailored-Controller/blob/main/Build/Controller/CoinChaserController.html}{HTML file} and added it to a \href{https://github.com/SmartControllerJS/Coin-Chaser-Tailored-Controller}{cloned game repository}, hosting it on GitHub Pages so that the QR code generated can direct users to that page after a connection is established. Within this file, we imported the SmartController package to assign this controller as an extension of the BaseController class and manage connections with the SmartPhoneController class. This was done by assigning "\textit{var phone = new smartcontroller.SmartPhoneController();}" and then sending command inputs from the controller to the phone with the \textit{phone.sendMessage(message);} command. For more information on how to create a controller script, refer to \href{https://smartcontrollerjs.github.io/SmartController/smartcontroller.html}{this documentation}. \\

\begin{figure}[b]
        \centering
        \begin{subfigure}[b]{0.175\textwidth} 
            \centering
            \includegraphics[width=\linewidth]{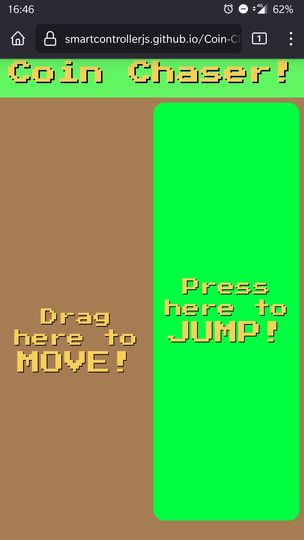}
             \caption[he Coin Chaser Controller in portrait mode.]%
            {{\small The Coin Chaser Controller in portrait mode.}} 
            \label{Portrait}
        \end{subfigure}
        \hfill

        \centering
        \begin{subfigure}[t]{0.3\textwidth}  
            \centering 
            \includegraphics[width=\textwidth]{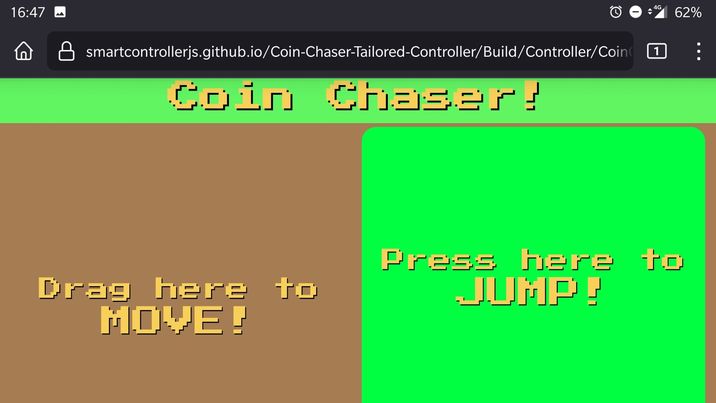}
            \caption[The Coin Chaser Controller in landscape mode.]%
            {{\small The Coin Chaser Controller in landscape mode.}}    
            \label{Landscape}
        \end{subfigure}
        \caption[ The tailored controller for Coin Chaser. ]
        {\small The tailored controller for Coin Chaser.} 
        \label{CoinChaserController}
\end{figure}

We designed and implemented this file with browser scripting, resulting in the controller layout seen in Figure \ref{CoinChaserController} and demonstrated online \href{https://smartcontrollerjs.github.io/Coin-Chaser-Tailored-Controller/}{here}. This controller features a joystick on the left-hand side column to control character movement, and a button on the right-hand side column to control character jumps. The reduction of the many buttons from the NES controller to just two inputs would hopefully mitigate unintentional button presses by users and reduce the number of glances from the game screen to the controller. Furthermore, users will not have to stretch their fingers to the center of the left column to move because the joystick dynamically appears where the user touches on the left column.\\

After designing the controller layout we connected it to the QR code generator in our main WebGL \href{https://github.com/SmartControllerJS/Coin-Chaser-Tailored-Controller/blob/main/Game/Assets/WebGlTemplates/WebsiteTemplate/index.html}{template browser script}, replacing the NES controller with our own. Finally, we modified our button input assignments to register the new controller joystick and jump button inputs. \\

\section{Limitations}
Some compatibility issues arose with Apple devices and the Safari browser, as well as with some brands of smartphones running Android OS. For example, the joystick controller relied on using an external library called \href{https://yoannmoi.net/nipplejs/}{Nipple.js} which had no support for Apple devices. To avoid this affecting the experiment, Samsung Android devices were supplied to participants when possible. In other cases, the smartphone could simply not connect via WebRTC and the source of this issue remains unknown. \\

While running our experiment on a University campus, we relied on the \href{https://eduroam.org/}{eduroam} wifi network which, at Glasgow University, was restricting the use of WebRTC. While WebRTC was chosen to keep our running costs to the minimum, it might be a limiting factor on some networks that might limit its usage. \\

\section{Discussion}

The user interviews revealed that the library is easy to use with tutorials available, well documented, and could be applied to replace mouse and keyboard input in browser games. \\

For Gotta Graduate, we used the Phaser framework and the SmartController library. Once both packages were imported, using the SmartController was a seamless experience upon following the instructions on \href{https://smartcontrollerjs.github.io/SmartController/gettingstarted.html}{SmartControllerJS}. The only issue that arose was understanding how to connect the generated QR code with the HTML of the webpage to ensure it was displayed correctly. Using the SmartController placed development focus on the core elements of the gameplay rather than the intricacies of using a new interaction method. Overall, we recommend Phaser as a game engine compatible with SmartController. In addition, JavaScript-based libraries and frameworks work effortlessly alongside the SmartController package and are highly recommended to allow the focus to be placed on the application rather than the interaction method. \\

The Joystick Controller helped create a smooth experience for participants playing Gotta Graduate. The dynamic, 360° movement allowed participants to focus on their gameplay rather than using the controller. This was an important aspect when choosing the controller as more buttons on the controller would lead to a higher learning curve and vice versa. However, development was more advanced with the joystick as it needed to be programmed to move in each direction using degrees ranging from 0 to 360. This led to problems with players repeatedly moving diagonally on the map of Gotta Graduate - an undesirable side effect. To cancel this effect, depending on the axis of movement, the opposite axis velocity would be set to 0, removing the accidental diagonal movement for players. \\

After observing numerous groups and solo participants playing Gotta Graduate, it became clear that in a public and social context, people playing in groups were seemingly more engaged than their solo counterparts. This was apparent after groups of players expressed more willingness to interact with the game. It became clear throughout the experiment that using the SmartController was a fluid enough interaction to allow players to feel comfortable using it within a public environment. As participants were already accustomed to using QR codes, there was little stopping them from instantly understanding how to play and interact with the game using the SmartController. Playing in groups often led to social learning taking place - when participants would help each other understand foreign concepts or tools. When a player understood the SmartController, they would often take the time to explain it to their fellow participants, allowing them to begin the game. This provides further evidence of the success of the SmartController as a means for interacting with the general public. \\

For Coin Chaser, once the communication between the browser script and Unity game was established, the SmartController was easy and simple to implement. The various demos on \href{https://smartcontrollerjs.github.io/SmartController/gettingstarted.html}{SmartControllerJS} effectively guided the methods to generate, and display, a QR code as well as to implement the selected NES controller. Observing the results obtained from the Coin Chaser evaluation, it can be seen that the SmartController library has the potential for wider deployment with Unity browser games. \\

The NES controller was very easy to use with Coin Chaser. It was a simple task to connect the buttons on the interface to the expected inputs, using them as replacements for buttons seen on normal webpages. Players were able to adapt quickly to the set control inputs, with experienced gamers adapting to the controls during the first game round and the less experienced players adapting during the second game round.

\section{Conclusion}

We have successfully implemented SmartControllerJS, a JavaScript library that turns users’ smartphones into controllers. Once the initial framework was in place, Gotta Graduate and Coin Chaser were developed, showing that the library can be used to develop applications quickly and with increased accessibility at zero cost. The statistics discussed previously demonstrate the seamless connection the library creates between two web pages, typically on a desktop and smartphone, and the fast communication that takes place between the web-application and controller. Participant feedback from two experiments provided evidence that the SmartController is easy to use, with 68.75\% of users in one of the experiments agreeing that the controller is intuitive. This can be attributed to the use of QR codes to establish a connection and the familiarity users have with touchscreen input. \\

All the initial aims of the project have been satisfied and through performing the two experiments we have provided evidence for each.

\begin{enumerate}
    \item As seen in both experiments, the SmartController is easily adapted to unique web-based projects. The documentation and support have greatly increased over time to a level at which issues can often be resolved individually by developers without the intervention of the original library creator. For users, it was observed that the SmartController provided a quick and easy-to-understand method for participants to interact with both games without installation or detailed instruction.
    \item Gotta Graduate, Coin Chaser, and the development of a custom controller were all created and usable cost-free.
    \item As participants played on their own devices, both computers and phones, it has been demonstrated that both games are easily distributable. The results from the experiments also proved that participants were comfortable with the method of connecting their phone to the application via the SmartController.
    \item The SmartController has enabled both single and multiplayer experiences as observed in the experiments.
    \item A dedicated \href{https://smartcontrollerjs.github.io/Controllers/stats.html}{statistics page} has been developed to test the message rate and ping of any controller.
    \item The SmartController is available on GitHub at \href{https://github.com/SmartControllerJS}{SmartControllerJS} . In addition, the package has been deployed on NPM - \href{https://www.npmjs.com/package/smartcontroller}{npmjs} - and UNPKG - \href{https://unpkg.com/browse/smartcontroller@3.2.7/}{unpkg}. 
\end{enumerate}

\section{Future Work}
SmartController should prioritise controllers compatible with all devices and browsers. For example, the joystick controller should be remade without using the Nipple.js library to provide a joystick that works correctly on all devices. The documentation on how to implement a custom controller should also be improved, with step-by-step instructions. In the same vein, we plan to cover how to interface SmartController with various game frameworks, such as Phaser and Unity, and will include a dedicated section on multiplayer capabilities. \\

Our aim is to expand the number and type of controllers available. For instance, modern phones have an abundance of technology that is currently untapped such as cameras, gyroscopes, proximity sensors, and more. One could easily imagine detecting hand motion using image recognition from a camera sensor and using it to manipulate web-applications. This would complement the current task of the SmartController and provide a suitable alternative to using a smartphone’s touch capabilities. \\

Beyond the library itself, we believe the SmartController concept could be expanded in two directions. \\

The first direction is to enable A/B testing and the hyper personalisation of controllers. Applying best practices from web development to the development of controllers might be a promising approach. For example, players of mainstream gaming consoles, such as the Playstation or Xbox, experience games using a unique physical controller. However, some games might benefit from extremely customised controllers and the quality of these controllers could be decided by experimentation. A/B testing is one way to do so in a data driven way. Another approach could be to cluster users and ship a different controller per user, or, even further, evolve or optimise a controller to the needs and preferences of a specific user. We have experimented at the extreme end of this space with a PIN entering interface, called IFTTT-PIN \cite{grizou_2022}, that allows users to decide which button to use for which action on the fly, without informing the web-app, see \href{https://jgrizou.github.io/IFTT-PIN/}{IFFT-PIN}. Combining IFTT-PIN concepts with SmartControllerJS capabilities and A/B testing principles is a promising way towards highly customisable user interfaces. \\

The second direction is to expand the SmartControllert concept to physical objects. Could we scan a QR code on our TV and control it with our phone? Could children's toys be controlled that way? Could we offer the user the ability to reprogram those physical objects along with the interaction we propose with SmartController? In the spirit of SmartController this should be accomplished via a web page requiring no installation, no server cost, and be easy to scale. \\

We are exploring this using \href{https://www.espruino.com/}{Espruino}, a framework to run JavaScript on a microcontroller. With web-Bluetooth capabilities, it is possible to connect and reprogram a device directly from your browser and on your phone. We developed the Espruino Remote Uploader tool to facilitate experimentation, for which a demo can be found at \href{https://cmurray95.github.io/espruino-remote-uploader/demos/puck/colour.html}{https://cmurray95.github.io/espruino-remote-uploader/demos/puck/colour.html}. The user can connect to the device, upload custom code, and interact with it via its smartphone. Pushing this concept further, we developed a self-driving robot that is controlled directly by a smartphone from a webpage. You can find this project at \href{https://github.com/Kirstin813/L4-Individual-Project}{https://github.com/Kirstin813/L4-Individual-Project} which includes many video demonstrations. 

If you have any questions, you can raise an issue on \href{https://github.com/SmartControllerJS/SmartController}{SmartControllerJS GitHub} or directly email the authors.

\newpage

\section{Table of demos}

\begin{center}
\begin{minipage}{\linewidth}
    \centering
\renewcommand{\arraystretch}{1.5}
\begin{tabularx}{\textwidth} { 
  | >{\centering\arraybackslash}X |
 }
\hline
Interactive Experiments: \\
 \hline
 \href{https://fraser-dempster.github.io/l4-project-interactive-game/}{Gotta Graduate} \\
  \hline
 \href{https://smartcontrollerjs.github.io/Coin-Chaser/}{Coin Chaser} \\
 \hline
  \href{https://smartcontrollerjs.github.io/Coin-Chaser-Tailored-Controller/}{Coin Chaser with the Tailored Controller} \\
 \hline
SmartControllerJS demos: \\
 \hline
 \href{https://emmapoliakova.github.io/WebRTCSmartphoneController/demo/tinyPlatformer/index.html}{Platformer game} \\
  \hline
 \href{https://emmapoliakova.github.io/WebRTCSmartphoneController/demo/3dRacing.html}{3D racing} \\
  \hline
 \href{https://emmapoliakova.github.io/WebRTCSmartphoneController/physics/physicsDemoV3.html}{Physics simulator} \\
 \hline
\href{https://emmapoliakova.github.io/WebRTCSmartphoneController/physics/physicsDemoV4.html}{Physics simulator - multiplayer} \\
 \hline
\href{https://emmapoliakova.github.io/WebRTCSmartphoneController/handtracking/receiveVideo.html}{Hand tracking} \\
\hline
\href{https://smartcontrollerjs.github.io/Controllers/touchpad-receive.html}{Multiplayer touchpad} \\
\hline
\href{https://smartcontrollerjs.github.io/Controllers/controller-receive.html}{Nes controller arrows} \\
\hline
\href{https://smartcontrollerjs.github.io/Controllers/joystick-receive.html}{Joystick} \\
\hline
\href{https://smartcontrollerjs.github.io/Controllers/stats.html}{Statistics page} \\
\hline
\href{https://smartcontrollerjs.github.io/SmartController/}{Documentation} \\
\hline
\end{tabularx}
\captionof{table}{Table of all available experiments and demos. }
\label{table:tableofdemos} 
\end{minipage}
\end{center}

\newpage

\section{Bibliography}
\bibliography{References.bib}

\end{document}